# Frequency and Amplitude Optimizations for Magnetic Particle Spectroscopy Applications

Vinit Kumar Chugh[a], Arturo di Girolamo[a], Venkatramana D. Krishna[b], Kai Wu[a,#,*], Maxim C-J Cheeran[b,*], and Jian-Ping Wang[a,*]

[a]Department of Electrical and Computer Engineering, University of Minnesota, Minneapolis, MN 55455, United States

[b]Department of Veterinary Population Medicine, University of Minnesota, St. Paul, MN 55108, United States

*E-mails: kai.wu@ttu.edu (K. W.), cheeran@umn.edu (M. C-J. C.), and jpwang@umn.edu (J.-P. W.)

#*current address:* Department of Electrical and Computer Engineering, Texas Tech University, Lubbock, TX 79409, United States

**Abstract**

Nowadays, there is a growing interest in the field of magnetic particle spectroscopy (MPS)-based bioassays. MPS monitors the dynamic magnetic response of surface-functionalized magnetic nanoparticles (MNPs) upon excitation by an alternating magnetic field (AMF) to detect various target analytes. This technology has flourished in the past decade due to its low cost, low background magnetic noise interference from biomatrix, and fast response time. A large number of MPS variants have been reported by different groups around the world, with applications ranging from disease diagnosis to foodborne pathogen detection, and virus detection. However, there is an urgent need for guidance on how to optimize the sensitivity of MPS detection by choosing different types of MNPs, AMF modalities, and MPS assay strategies (i.e., volume- and surface-based assays). In this work, we systematically study the effect of AMF frequencies and amplitudes on the responses of single- and multi-core MNPs under two extreme conditions, namely, the bound and unbound states. Our results show that some modalities such as dual-frequency MPS utilizing multicore MNPs are more suitable for surface-based bioassay applications, whereas, single-frequency MPS systems using single- or multi-core MNPs are better suited for volumetric bioassay applications. Furthermore, the bioassay sensitivities for these modalities can be further improved by careful selection of AMF frequencies and amplitudes.

**Keywords:** *Magnetic particle spectroscopy, magnetic particle imaging, magnetic nanoparticle, bioassay, disease diagnosis, virus detection*

## 1. INTRODUCTION

Magnetic particle spectroscopy (MPS) is a technology derived from the non-invasive tomographic technique, magnetic particle imaging (MPI)[1,2]. The idea of MPS was firstly conceived by Krause *et al.* and Nikitin *et al.* in 2006, respectively[3,4]. MPS detects the dynamic magnetic responses of magnetic nanoparticles (MNPs) upon the application of alternating magnetic fields (AMFs), namely, the driving fields. Over the last decade, MPS has been reported as a powerful tool for detecting blood viscosity and blood clot[5,6], foodborne pathogens and toxins[7–10], animal (e.g., Hepatitis B, H1N1, SARS-CoV-2, etc.)[11–17] and plant viruses (e.g., Grapevine fanleaf virus, Potato virus X, etc.)[18], hormones, cytokines and other signaling molecules from body fluid[19,20]. Nowadays, there have emerged a variety of MPS platforms such as the single- and dual-frequency AMF implementations (categorized by the number of AMFs), the single- and multi-core MNP-based assays (categorized by the structure of MNPs), and the volumetric- and surface-based assay strategies (categorized by the major contributing relaxation mechanisms)[12,14,21–24].

The MPS volumetric systems mainly rely on detecting subtle changes in the binding state of the MNP (binding to target analytes or cross-linked by the presence of target analytes), which results in a difference in Brownian relaxation time and changes the magnetic response. Both single- and multi-core MNPs combined with either single- or dual-frequency AMFs have been reported on MPS volumetric systems. For example, 80 nm multi-core MNPs and 30 nm single-core MNPs have been independently reported in the MPS volumetric system for the detection of SARS-CoV-2 proteins[12,14]. Multi-core MNPs combined with single-frequency excitation field-driven MPS volumetric systems have also been widely reported for detecting a variety of disease biomarkers such as thrombin, inflammation, and infection biomarkers, etc[20,23,25]. On the other hand, in MPS surface-based assay systems, the multi-core MNPs driven by dual-frequency AMFs are widely reported for bioassay applications[8,18,26–29]. In this scenario, MNPs are captured on a solid substrate such as a nonmagnetic porous filter or nitrocellulose membrane through the specific antibody – antigen – antibody sandwich assay structure. The dynamic magnetic responses are due to the Néel relaxation of magnetic moments in these captured MNPs.

However, there is a lack of systematic studies on how to choose the right type of MNPs (single- or multi-core structures) for certain MPS-based bioassay applications. Furthermore, guidelines for selecting optimal AMF frequencies and amplitudes for certain types of MNPs to achieve maximum bioassay sensitivity are urgently needed. In this work, we show an example of comparing the dynamic magnetic responses of two types of MNPs (i.e., 30 nm single-core SHS30 MNP and 50 nm multi-core SuperMag50 MNP) under single- and dual-frequency AMFs. Two extreme bioassay conditions are designed, namely, the unbound MNPs suspended in liquid and the fully cross-linked MNP clusters in liquid. For the latter case, an excess number of biotinylated antibodies are added to ensure that all MNPs are bound, and the Brownian relaxation is fully blocked. This condition also favors the study of surface-based MPS assay since under this condition only Néel relaxation is governing the relaxation dynamics of MNP magnetization. Thus, the work reported herein is tuning the three variables: MNP structures,

AMF, and the MPS assay strategy intending to provide peers with insights on designing optimal settings for MPS-based bioassays.

The results show that the single-frequency MPS modality utilizing both the single- and multi-core MNPs is suitable for volumetric bioassay applications. The sensitivity of this methodology can be further improved by utilizing higher excitation frequencies and amplitudes as both promote the Brownian relaxation mechanism of nanoparticles. It was also observed that in dual-frequency MPS systems, utilizing multi-core MNPs and operating at higher excitation frequencies are optimal for surface-based bioassay applications, as Néel relaxation is the dominant relaxation mechanism. It was also noted that the sensitivity of such surface-based bioassay systems could be further optimized by a careful selection of high- and low-frequency excitation field amplitudes.

## 2. MATERIALS AND METHODS

### 2.1. Materials

The SHS30 MNPs are streptavidin-coated 30 nm single-core iron oxide nanoparticles and SuperMag50 are 50 nm multi-core iron oxide nanoparticles conjugated with streptavidin, both at concentrations of 1 mg/mL were obtained from Ocean NanoTech. Biotinylated antibody goat anti-mouse IFN-γ with an antibody concentration of 0.4 µg/mL (2.67 nM) was obtained from R&D systems (ELISA kit product number DY485-05, part number 840123). Phosphate buffered saline (PBS) was purchased from Genesee Scientific.

### 2.2. Static Magnetic Property Characterization

Static magnetic property characterization of the MNP samples was performed using the PPMS DynaCool system from Quantum Design. MNP samples were prepared by air drying 10 µL of 1 mg/ml MNP solutions on filter paper. Measurements were performed at 300 K and by varying the quasi-static field in ±5000 Oe or ±500 Oe ranges. Further information regarding the measured static properties is provided in Supporting Information S1.

### 2.3. Modified MPS System

A modified version of our benchtop MPS system[11] added with a capacitor bank and a current feedback path was utilized for the optimization study experiments. Figure 1 shows a schematic representation of the modified MPS system which can be broken down into five different components: (1) a PC with pre-installed LabVIEW software to enable the system control; (2) A DAQ unit (NI USB-6289) for both the generation of sinusoids for excitation fields and subsequently the real-time collection of pick-up coil signal and passive current feedback information; (3) Power amplifiers to amplify the high and low-frequency excitation signals; (4) Switchable capacitor bank constituting of individual resonating capacitors for different frequencies; (5) Coil arrangement constituting of two excitation coils to generate single- and dual-frequency magnetic fields along with a balanced set of pick-up coils. An electrical readout is generated from pick-up coils based on the magnetization response of MNPs following Faraday's law of induction ($\varepsilon = -N\Delta\varphi/\Delta t$). Fast Fourier transform (FFT) is then applied to the captured electrical readout to analyze the harmonic response in different excitation scenarios. The presence of a current feedback loop enabled us to sense and control the magnetic field amplitude precisely while varying the excitation field frequencies. Resonating capacitors allowed us to push the operational frequencies for both the high frequency ($f_H$) and the low frequency ($f_L$) excitation field signal to higher values whilst ensuring low reactive losses and thus maintaining a reasonably applied voltage level. More on resonance effects and how it enables a high-frequency operation has been covered in Supporting Information S2.

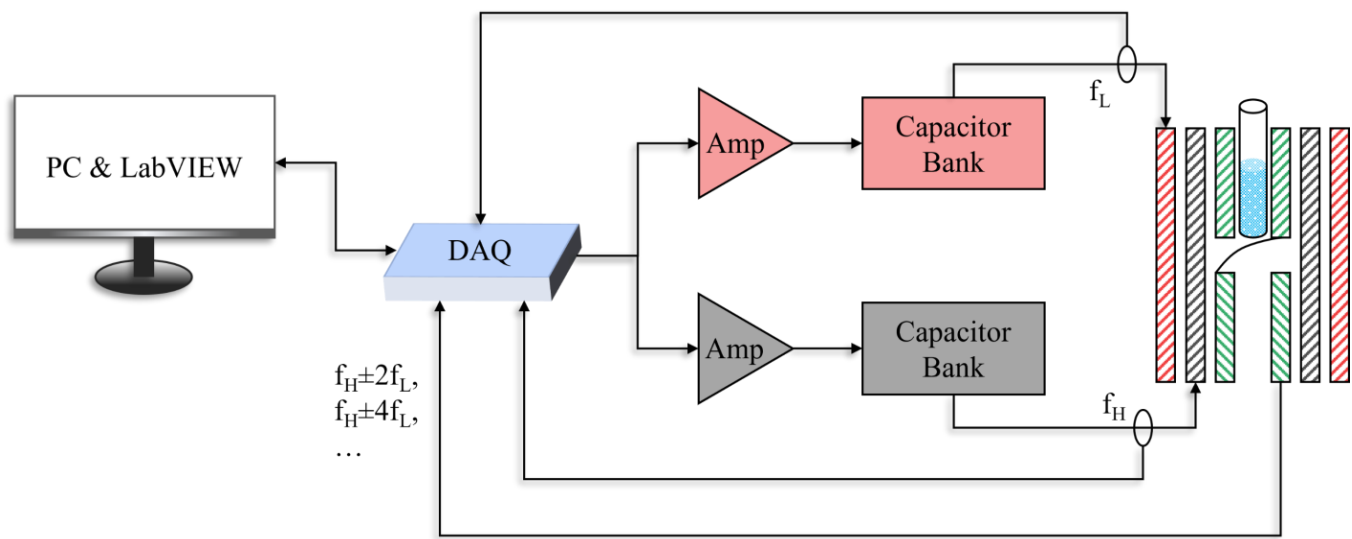

Figure 1. Schematic depiction of the modified benchtop MPS system with resonating capacitor bank for low and high-frequency driving coils.

### 2.4. Experimental Samples

Our experimental plans consisted of 2 sample groups utilizing SHS30 and SuperMag50 MNPs. Two samples were prepared for each group to simulate the unbound and bound states of the respective nanoparticles as listed in Table 1. It should be noted that although the weight concentrations of SHS30 and SuperMag50 MNPs are identical (1 mg/mL), the molar concentrations are different. Excess amounts of biotinylated antibodies were used to cause complete sedimentation of magnetic nanoparticles as observed in Figure 2(a). For the bound state samples, the SHS30 or SuperMag50 MNPs were mixed with biotinylated antibodies and incubated at 4°C overnight. The consequent cluster formation due to streptavidin-biotin binding events for single- and multicore MNPs have been depicted in Figure 2(b) and (c) respectively.

Table 1. Sample design utilizing single- and multicore nanoparticles for the MPS optimization study.

| Sample Index | MNPs used | MNP weight amount/vial | MNP molar amount/ vial | Biological Matrix | Antibody amount |
|---|---|---|---|---|---|
| I | SHS30 | 0.016 mg | 544 fmole | PBS | NA |
| II | | 0.016 mg | 544 fmole | Biotinylated Antibodies | 170.9 fmole |
| III | SuperMag50 | 0.04 mg | 40 fmole | PBS | NA |
| IV | | 0.04 mg | 40 fmole | Biotinylated Antibodies | 106.8 fmole |

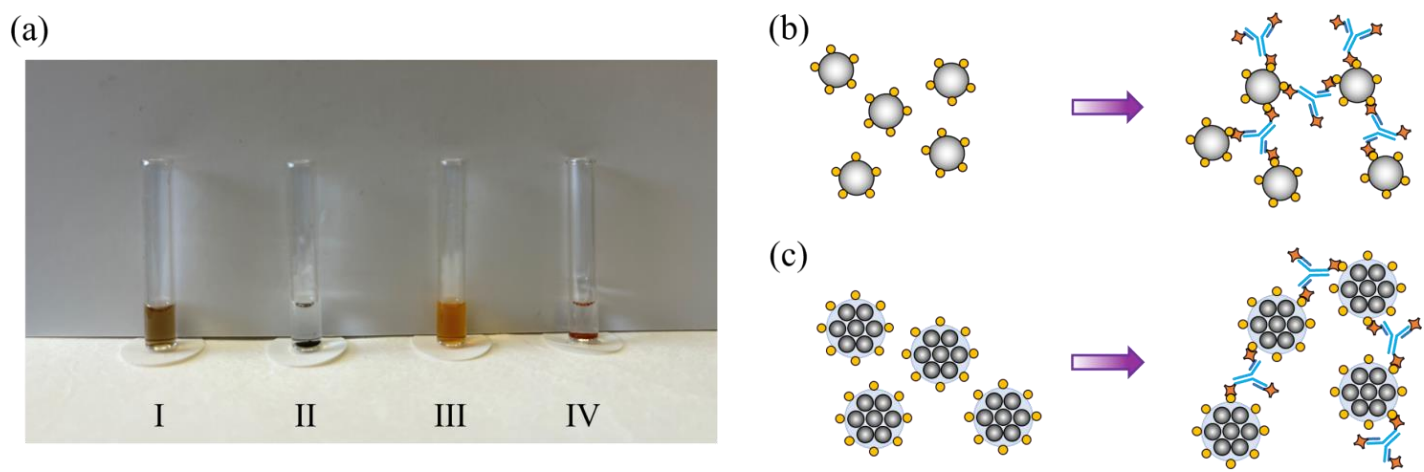

Figure 2. (a) Photographs of the prepared single-core SHS30 (vials I and II) and multicore SuperMag50 (vials III and IV) samples. Complete sedimentation of the MNPs was observed from vials II and IV where excess amounts of biotinylated antibodies were added. Schematic presentation of the cluster formation of single-core and multicore MNPs presented respectively in (b) and (c).

## 3. RESULTS AND DISCUSSIONS

### 3.1. Dynamic Magnetic Property Characterization of MNP Samples

The dynamic magnetization response of the MNP samples was studied under the application of a single-frequency AMF with an amplitude of 250 Oe and a frequency of 130 Hz. Dynamic property characterization was performed using the same modified benchtop scheme as described in Figure 1 with only the excitation of the low-frequency field coil. The magnetization response of the MNPs results in corresponding voltage generation in the balanced pickup coil following Faraday's law of induction. This voltage response of the MNPs was then captured at a sampling rate of 100 ksps using the data acquisition channel of DAQ. The real-time voltage response along with the information on the applied magnetic field was then utilized to plot the dynamic magnetization-field (M-H) response curve following the methodology presented in Supporting Information S3. We do acknowledge that the magnetization response captured would look differently if captured under a different excitation frequency and or magnetic field amplitude[30].

Dynamic magnetization responses of both single-core SHS30 and multi-core SuperMag50 MNPs in their bound and unbound states are presented in Figure 3(a) and (b), respectively. In both plots, the response has been normalized to the highest magnetization values observed, namely, the unbound state of respective MNPs. A clear difference in the AC M-H loop can be observed from MNPs of the bound and unbound states in terms of increased coercivity and reduced remnant magnetization values. This change thus explains the reduction in harmonic amplitude of captured MPS spectra for bound state nanoparticles as compared to the respective unbound state.

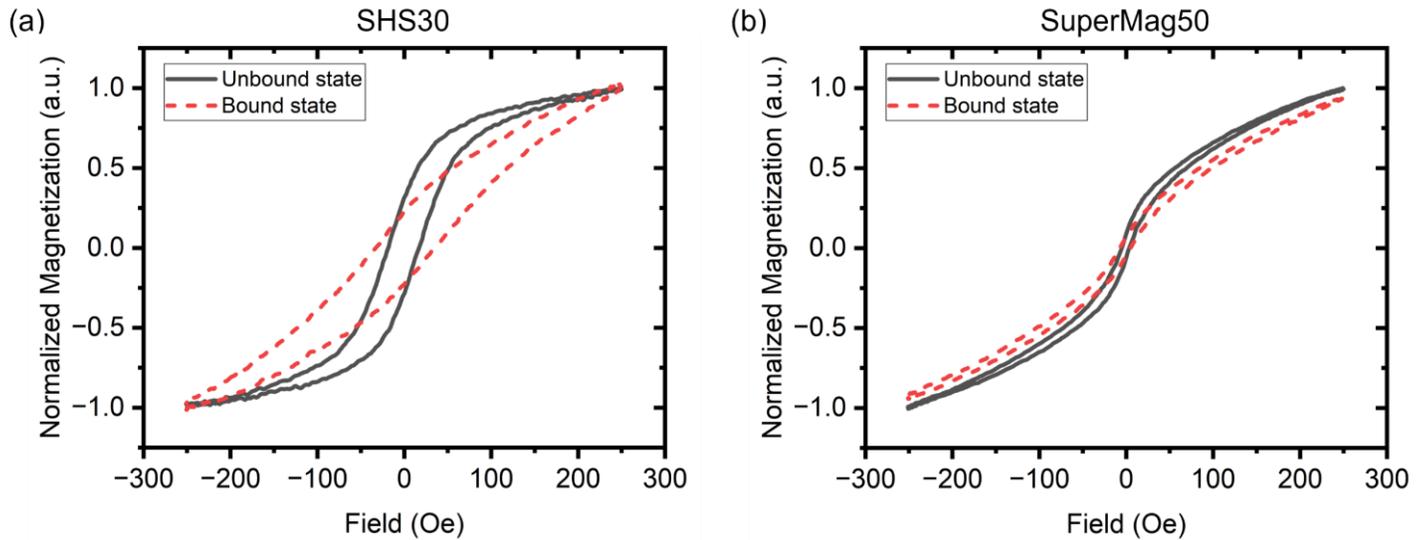

Figure 3. The AC M-H curves of (a) SHS30 and (b) SuperMag50 MNPs in bound and unbound states.

With this initial characterization of MNP samples done, further experiments were carried out to study the dependence of MPS spectra performance on the excitation field frequencies and amplitudes. The systematic experiments carried out for both single- and dual-frequency MPS systems for frequency and field amplitude optimizations have been presented with detailed analysis in the following Sections 3.2 and 3.3 respectively.

### 3.2. Frequency Optimization

Frequency optimization experiments were performed to study the dependence of harmonic spectra of MNPs on the excitation field frequencies in an MPS system. Experiments were designed to include frequency variations for both single- and dual-frequency MPS methodologies. For studying single-frequency MPS modality, the voltage was applied only to the low-frequency coils from the system described in Figure 1 keeping the coil arrangements the same between experiments. Whereas for the dual-frequency experiments, both the low- and high-frequency coils were applied with a sinusoidal excitation signal. We will investigate the frequency optimization results for single- and dual-frequency systems in Sections 3.2.1 and 3.2.2, respectively.

### 3.2.1 Single Frequency System

For the frequency optimization studies on the single-frequency system, MPS harmonic spectra were recorded with excitation of the low-frequency ($f_L$) coil at various frequency values while maintaining a constant field amplitude of 250 Oe. $f_L$ was varied between 50 Hz, 130 Hz, 285 Hz, 620 Hz, and 1.38 kHz. The choice of these peculiar frequency numbers was due to the limited availability of unique capacitance values for the resonant capacitors. For each excitation frequency, we recorded the MPS spectrum and analyzed the changes in the 3$^{rd}$ and 5$^{th}$ harmonics. Figure 4(a) shows the frequency dependence of the 3$^{rd}$ harmonic response of bound and unbound states for SHS30 and SuperMag50 MNPs. The results corresponding to the 5$^{th}$ harmonic variations are depicted in Figure 4(b). We also calculated the % drop in harmonic signal for the two states (defined as % $Drop\,(\Delta) = \frac{A_{Bound\,State} - A_{Unbound\,State}}{A_{Bound\,State}} \times 100$, where A stands for the harmonic amplitude) and the results have been presented in Figure 4(c). From these readings, several key observations can be made as noted. Firstly, both the 3$^{rd}$ and the 5$^{th}$ harmonic amplitudes increase with an increment of the excitation frequency $f_L$ for SHS30 as well as SuperMag50 MNPs. Secondly, Δ increases linearly for the single-core SHS30 MNPs and almost exponentially for the multicore SuperMag50 MNPs for the frequency range tested. Thirdly, Δ is higher for the SHS30 MNPs as compared with the SuperMag50 in the observed frequency values. Δ can also be understood as the contribution of the Brownian relaxation mechanism to the harmonic response of an MNP, as Δ is the reduction in signal from a state allowing both Néel and Brownian relaxation simultaneously to a state allowing only Néel relaxation process. This definition of Δ would be helpful in our further discussions. It should also be noted that the observations made above are valid for the frequency range mentioned previously.

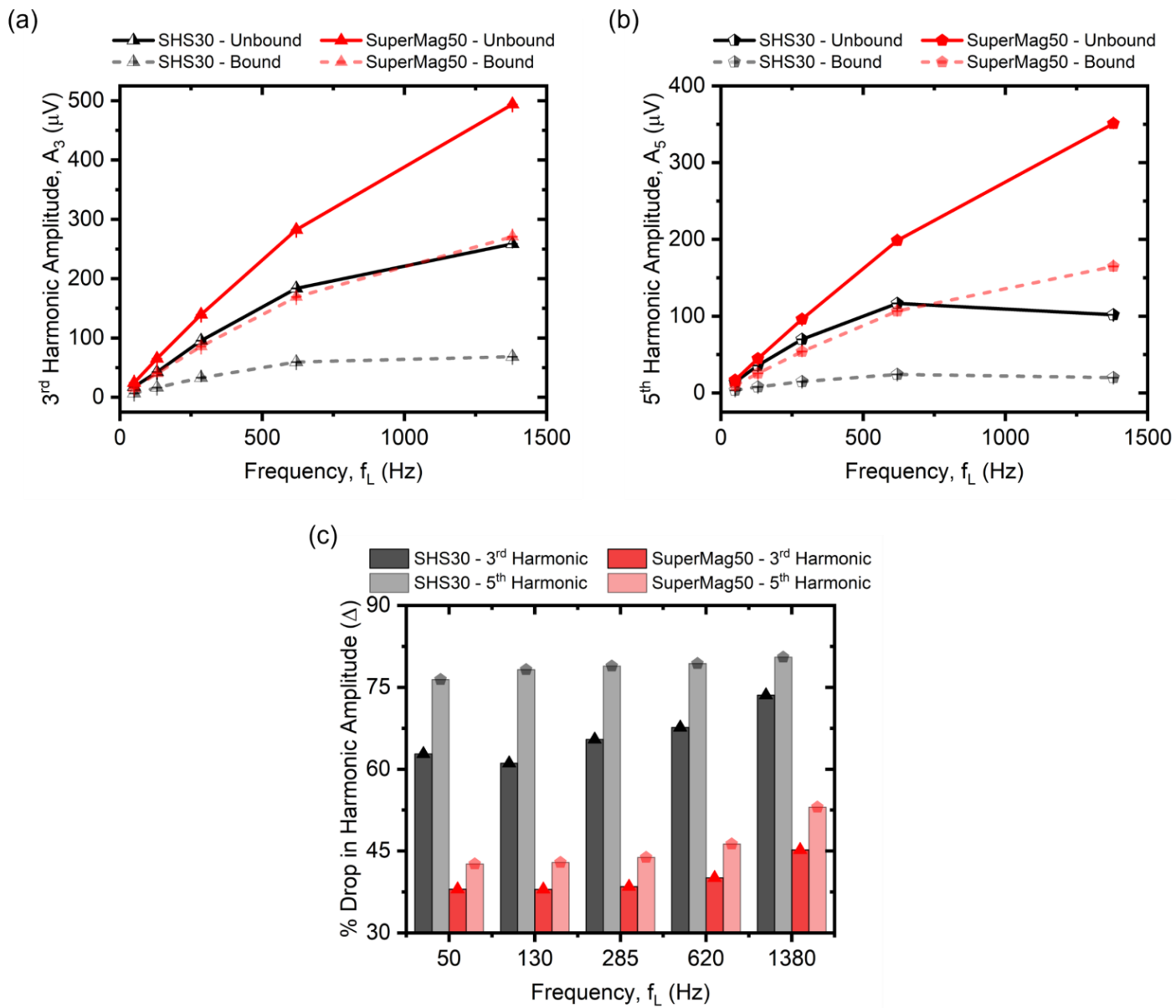

Figure 4. Frequency optimization results for the single-frequency system. (a) and (b) are the 3$^{rd}$ and the 5$^{th}$ harmonic amplitudes recorded from SHS30 and SuperMag50 MNPs of bound and unbound states. (c) is the calculated % drop in harmonic signal for both types of MNPs.

### 3.2.2 Multiplexed Frequency System

For the multiplexed (dual-) frequency investigations, both low-frequency ($f_L$) and high-frequency ($f_H$) excitation fields were applied to MNPs. For this portion of the study, the goal was to examine the effect of $f_H$ variation whilst keeping all other parameters constant to limit the scope. Thus, to achieve this objective, the following experimental design was followed: (a) Low-frequency excitation signal was maintained at a constant operational frequency of $f_L$= 50 Hz and a constant field amplitude of 250 Oe; (b) amplitude of high-frequency field was maintained at 25 Oe whilst varying the excitation frequency $f_H$ between 1 kHz and 27 kHz. With these experimental designs maintained, the spectral response of bound and unbound MNP samples was recorded for the varying $f_H$ values. The observed frequency dependence of the 3$^{rd}$ and the 5$^{th}$ harmonic amplitudes of MNPs has been summarized in Figure 5(a) and (b). The signal drop in respective harmonic amplitude, Δ has also been presented in Figure 5(c). Key takeaways from these observations were as follows: Firstly, from all the samples, both the 3$^{rd}$ and the 5$^{th}$ harmonic amplitudes increase with the increasing $f_H$. Secondly, Δ decreases linearly for both the 3$^{rd}$ and the 5$^{th}$ harmonics of single-core SHS30 MNPs with an increment in the excitation frequency, $f_H$. Thirdly, Δ drops consistently for the multicore SuperMag50 MNPs reaching a baseline saturation level around $f_H$=14 kHz. And lastly, the observed drop was persistently larger for harmonics of SHS30 MNPs when compared to the drop in respective harmonics of SuperMag50 MNPs for the frequency range in observation. We will once again use these observations in further sections to conclude the optimal operating conditions for bioassay applications.

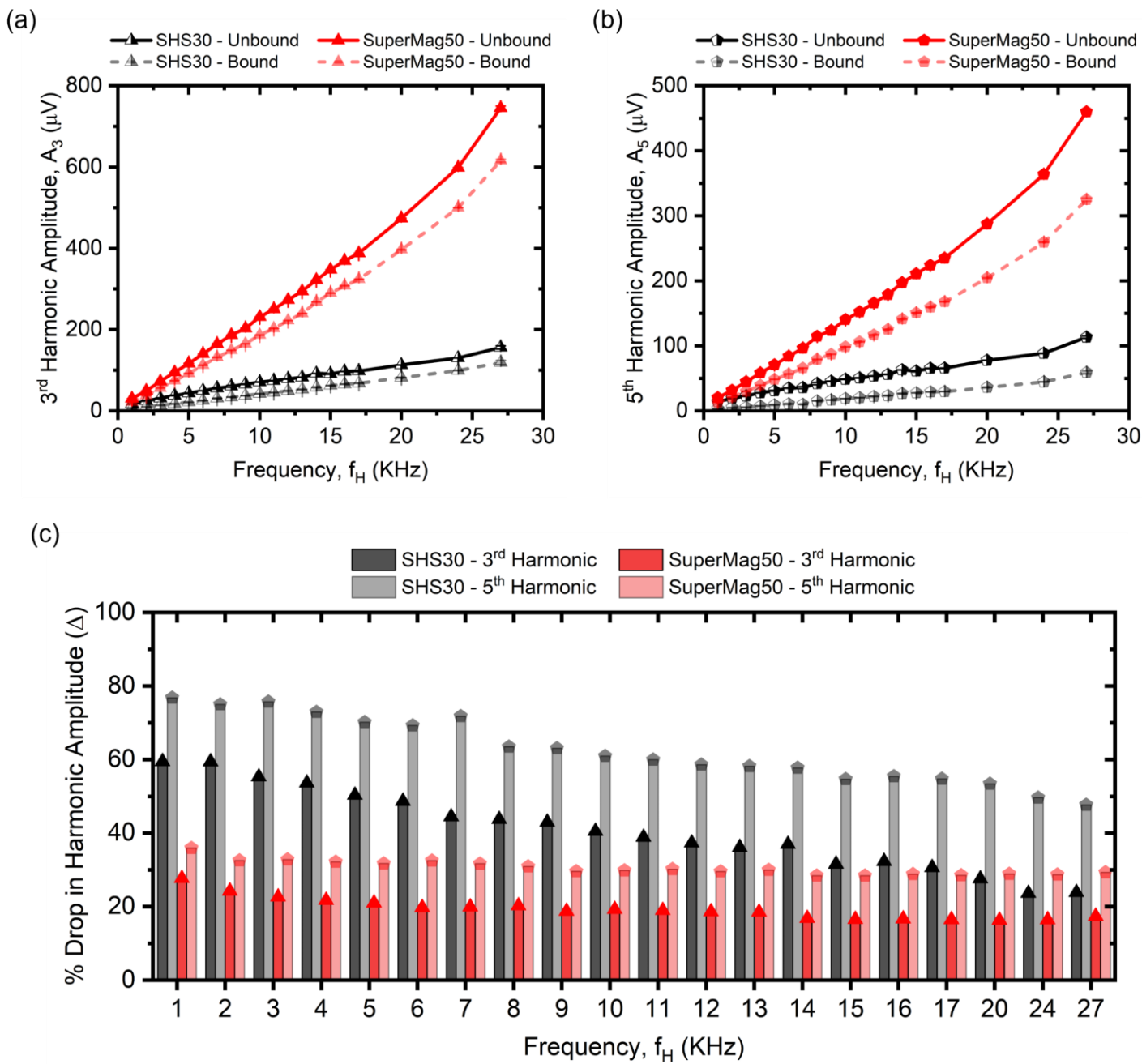

Figure 5. Frequency optimization results for the dual-frequency system. (a) and (b) are the 3$^{rd}$ and the 5$^{th}$ harmonic amplitudes recorded from SHS30 and SuperMag50 MNPs of bound and unbound states. (c) is the calculated % drop in harmonic signal for both types of MNPs.

### 3.3. Amplitude Optimization

Amplitude optimization studies were performed on the MNP samples by keeping the excitation frequencies, i.e., $f_L$ and $f_H$ constant whilst changing only the applied field amplitude. For a single-frequency MPS system, only the $f_L$ excitation field was applied and its amplitude $A_L$ was varied. For a dual-frequency MPS setup, both $f_L$ and $f_H$ fields were applied and their amplitudes, i.e., $A_L$ and $A_H$ were swept. Detailed study setup and results have been presented in the subsections.

### 3.3.1 Single Frequency System

The amplitude optimization study for single-frequency setup was carried out by keeping the field frequency, $f_L$ constant at 620 Hz and varying its amplitude $A_L$ between 31.25, 62.5, 125, and 250 Oe. MPS spectra were recorded from all 4 samples, using each of the prior mentioned excitation conditions. Hence in doing so, the dependence of both the 3rd and the 5th harmonic amplitudes on the excitation field amplitude was studied. Amplitudes of the 3rd and the 5th harmonics under varying excitation fields were summarized in Figure 6(a) and Figure 6(b), respectively. Similar to the frequency optimization studies, the drop in the respective harmonic amplitude was also calculated and has been plotted in Figure 6(c). The key observations from this part of the study are as stated. Firstly, the 3rd and the 5th harmonic amplitudes increase steadily for both bound and unbound states of the two types of MNPs with an increment in applied field amplitude. Secondly, the drop in harmonic amplitude from bound to unbound states, $\Delta$, increases almost linearly with a decrement in $A_L$. Thirdly, $\Delta$ for the 5th harmonic is decidedly larger than that of the 3rd harmonic. One abnormal reading of SHS30 MNPs for $A_L$ = 31.25 Oe can be explained by the presence of small amplitudes of harmonic magnitude for this case, and the corresponding instrumentation recording error causing this issue. Fourthly, $\Delta$ for SHS30 is observed to be larger than that for the SuperMag50 nanoparticles. These amplitude dependence results are in accordance with the theoretical models[31] for excitation field amplitude dependencies.

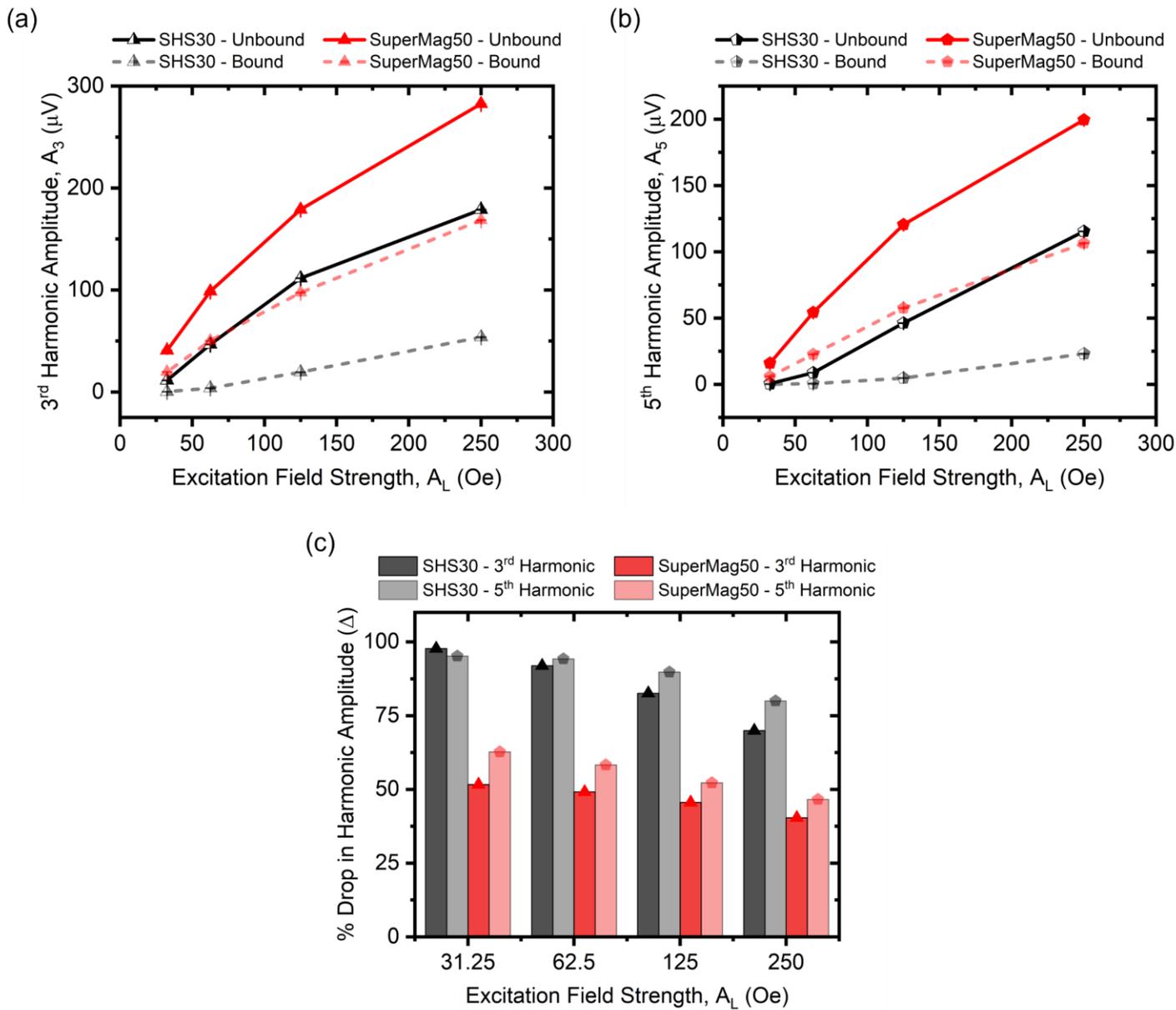

Figure 6. Amplitude optimization results for the single-frequency system. (a) and (b) are the 3$^{rd}$ and the 5$^{th}$ harmonic amplitudes recorded from SHS30 and SuperMag50 MNPs of bound and unbound states. (c) is the calculated % drop in harmonic signal for both types of MNPs.

### 3.3.2 Multiplexed Frequency System

The following experimental design was followed for amplitude optimization studies using a multiplexed (dual-) frequency system: (1) Field excitation frequencies, $f_L$ and $f_H$ were kept constant at 50 Hz and 5 kHz respectively; (2) Amplitude of the low-frequency field $A_L$ was varied between 31.25, 62.5, 125, and 250 Oe; (3) For each magnitude of the low frequency applied, the high-frequency field amplitude $A_H$ was varied between 2.78, 8.33, 16.67, and 25 Oe and the MPS spectra were collected for all of the resulting 16 unique excitation scenarios. The 3$^{rd}$ and the 5$^{th}$ harmonic amplitude results from this part of the study have been presented in Figure 7(a) and (b).

The resultant harmonic drop (Δ) from the unbound state of MNPs to the maximum binding state has also been depicted in Figure 7(c).

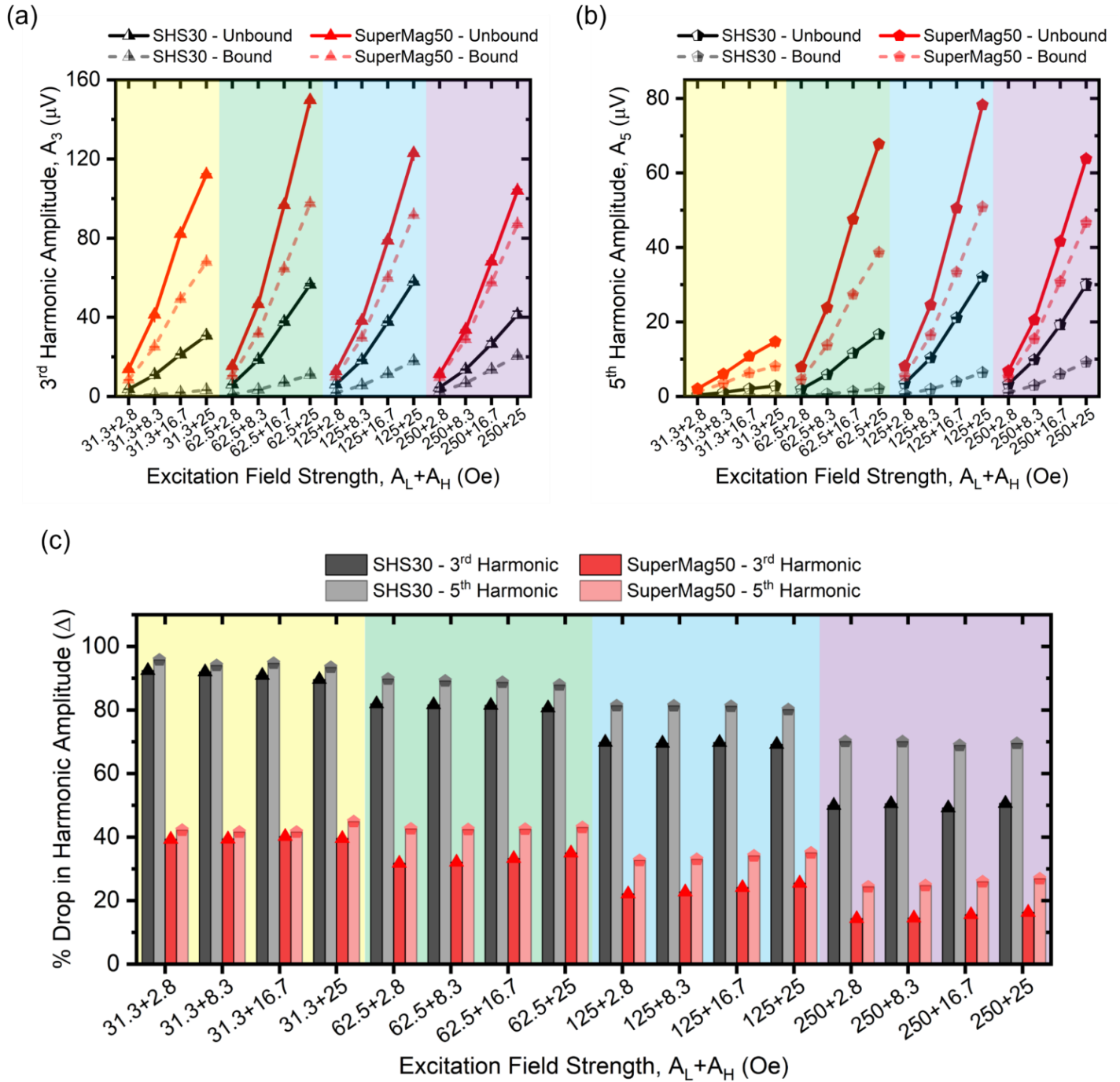

Figure 7. Amplitude optimization results for the dual-frequency system. (a) and (b) are the 3$^{rd}$ and the 5$^{th}$ harmonic amplitudes recorded from SHS30 and SuperMag50 MNPs of bound and unbound states. (c) is the calculated % drop in harmonic signal for both types of MNPs.

From these results obtained, the following insights can be drawn on the effect of excitation field magnitude on MNP harmonic spectra: (1) Reducing the $A_H$ amplitude leads to a monotonic reduction in both the 3$^{rd}$ and the 5$^{th}$ harmonic amplitudes for both the bound and the unbound MNP samples; (2) Reducing the $A_L$ does not lead to a monotonic reduction in respective harmonic amplitudes. The particular combinations producing the highest harmonic signal for SHS30 and SuperMag50 MNPs in different states have been captured in Table 2. These findings indicate that the choice of an excitation field for getting the highest harmonic amplitude is a function of all three factors, namely, (1) the choice of the nanoparticles, (2) the choice of MPS harmonic, and (3) the conjugation state (bound or unbound) of MNPs for a particular application, i.e., for surface-based bioassay applications, one might want to maximize the harmonic amplitude from the bound state of nanoparticles, whereas, for the volumetric bioassay applications, one might want to go with an excitation field magnitude choice maximizing the harmonic amplitude from unbound state and generating the highest Δ.

Table 2. The combinations of low- and high-frequency excitation field amplitudes, $A_L$ and $A_H$ generating the highest harmonic amplitude for SHS30 and SuperMag50 MNPs have been presented.

| MNPs | Rotational State | Harmonic | $A_L$ | $A_H$ |
|---|---|---|---|---|
| SHS30 | Unbound State | 3rd Harmonic | 125 Oe | 25 Oe |
| | | 5th Harmonic | 125 Oe | 25 Oe |
| | Bound State | 3rd Harmonic | 250 Oe | 25 Oe |
| | | 5th Harmonic | 250 Oe | 25 Oe |
| SuperMag50 | Unbound State | 3rd Harmonic | 62.5 Oe | 25 Oe |
| | | 5th Harmonic | 125 Oe | 25 Oe |
| | Bound State | 3rd Harmonic | 62.5 Oe | 25 Oe |
| | | 5th Harmonic | 125 Oe | 25 Oe |

Similarly, the following observations regarding the drop in harmonic amplitude can also be made: (1) Reducing the $A_H$ does not have a significant impact on Δ for either of the harmonics for both SHS30 and SuperMag50 nanoparticles, this can be observed from the relatively constant value of Δ for both the nanoparticles in Figure 7(c) when $A_L$ is kept constant; (2) A significant step increment in the Δ is observed for both the 3rd and 5th harmonics of SHS30 and SuperMag50 nanoparticles on reduction in the excitation field amplitude $A_L$; (3) Δ for single-core SHS30 is always greater than that observed for multicore SuperMag50; lastly, (4) Drop in the 5th harmonic in all scenarios for both the nanoparticles is found to be larger as compared to the respective drop in the 3rd harmonic. We will look into the particular implications of these results on the bioassay methodologies in the following sections.

### 3.4. Volumetric Bioassays

With all the frequency and amplitude dependencies as studied above for single-core SHS30 and the multicore SuperMag50 MNPs in both bound and unbound states, the key question of interest becomes what the optimum condition would be for the design of a bioassay system utilizing MPS or MPS-like methodologies. We will try to answer this key question to the best of our ability in this and the following section. In this section, we will discuss the implications for the design of a volumetric bioassay system and in Section 3.5, we will investigate the operational design implications for surface-based bioassay systems.

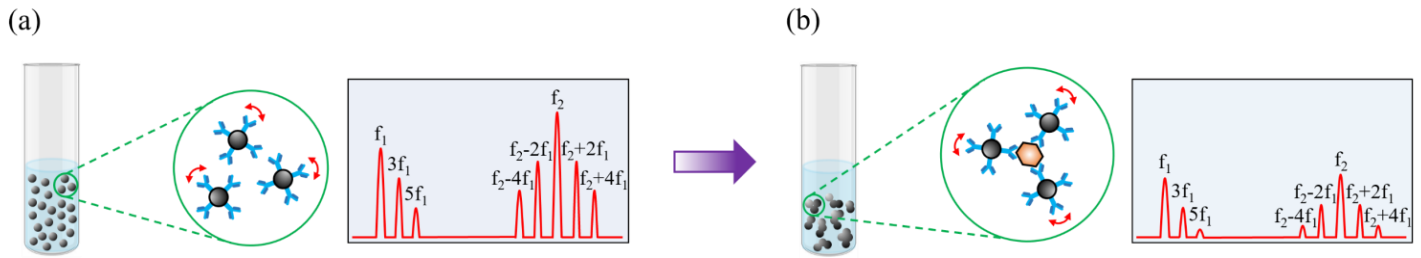

Figure 8. A schematic depiction of the harmonic change in a volumetric bioassay process for a dual-excitation field MPS system has been presented. (a) MNPs coated with polyclonal capture antibodies are free to rotate in the test volume; (b) Addition of biological sample constituting of target analyte causes cluster formation and increases hydrodynamic size. This increment in the hydrodynamic size leads to a higher phase lag in the Brownian relaxation process, causing a reduction in harmonic amplitude.

The schematic view of a volumetric bioassay methodology for a dual-frequency MPS system has been depicted in Figure 8. Free MNPs in a volumetric media can undergo both Néel and the Brownian relaxation process simultaneously[32,33] and hence have a larger harmonic response in MPS spectra. However, upon the addition of target analytes, the MNPs undergo clustering events thus restricting the Brownian relaxation which leads to a drop in the harmonic amplitude. Two important factors help improve the sensitivity of a volumetric bioassay MPS system: (1) Brownian dominant behavior of MNPs, what it means to our context is that a larger % drop (Δ) from unbound (simultaneous Brownian and Néel relaxation possible) to bound (only Néel relaxation possible) state benefits sensitivity; and (2) Higher harmonic amplitude from unbound MNPs, this has an indirect effect on sensitivity as it allows for use of smaller quantities of MNPs for bioassay testing, which has proven to improve the sensitivity significantly[34,35]. Because of this unique dependence, instead of giving general guidelines, we will limit ourselves to only a few categorical reviews that can help improve sensitivity.

For a single-frequency MPS system, increasing the excitation frequency $f_L$ positively improves both the harmonic amplitude from unbound MNPs and corresponding Δ, so the use of higher excitation frequencies can help with sensitivity improvement. Also, for the single-frequency MPS, increasing the excitation field amplitude increases the harmonic amplitude significantly, but also leads to a small reduction in Δ. Due to the relative scale of changes as observed from Figure 6, i.e., greater than 10× improvement in harmonic amplitude, and only around 25% reduction in corresponding Δ for the $A_L$ variation, an argument can be made that a larger excitation field is also beneficial for the overall sensitivity.

For a dual-frequency system, the excitation field amplitude combination ($A_L+A_H$) of 62.5 Oe + 25 Oe result in both the higher harmonic amplitude together with a significant improvement in Δ when compared to the excitation scenario of 250 Oe + 25 Oe excitation fields. This implies that a larger excitation signal is not necessarily the most optimal for bioassay applications. In the excitation frequency case, the scenario is much more complex as increasing the $f_H$ increases the harmonic amplitude significantly but also leads to a significant reduction in Δ (refer to Figure 5). Hence a coherent sensitivity argument cannot be made without conducting further tests to determine the limit of detection for each case.

### 3.5. Surface-based Bioassays

In surface-based bioassay methodologies, the MNPs are immobilized on a reaction surface following the sandwich bioassay mechanism. A schematic view of a sandwich bioassay scheme is depicted in Figure 9(a) along with the resultant harmonic spectra from immobilized MNPs. When immobilized MNPs can only undergo the Néel relaxation process once an excitation field is applied. This is also the case for the bound state of MNPs where complete sedimentation of nanoparticles can be observed (refer to Figure 2), i.e., samples II and IV, pertaining to strong Streptavidin-Biotin binding and excess amounts of biotinylated antibodies added. Hence, by looking specifically at the harmonic amplitudes of the bound state SHS30 and SuperMag50 nanoparticles, intuitive arguments can be made regarding the choice of nanoparticles and the MPS excitation methodology for optimum bioassay performance. It should also be pointed out here that this study was limited to the use of single-core SHS30 and multicore SuperMag50 nanoparticles, and a choice of a different set of MNPs can result in uniquely different observations.

Firstly, for the choice between a single- and a dual-frequency MPS methodology, the latter is more favorable. This can be argued by the observation that although increasing the excitation frequency in either of the methodologies leads to a monotonous increment in the harmonic amplitudes, in the case of single-frequency excitation, it also leads to a significant increment in Δ which is not the case observed for the dual-frequency excitation scenarios. Secondly, the use of multicore SuperMag50 MNPs is several orders better than the use of single-core SHS30 MNPs. The argument for this peculiar choice of MNPs can be understood from the findings presented in Figure 9, which shows the results for harmonic amplitude generated per micro-mole of the bound state of MNPs for both sets of nanoparticles. It can be noted that SuperMag50 generates a significantly larger harmonic response per micro-mole of nanoparticles. This can be attributed to two main reasons, (1) SuperMag50 MNPs show a strong Néel dominant relaxation mechanism for dual-excitation MPS systems as can be inferred from Figure 5(b); and (2) SuperMag50 MNPs are made of a cluster of smaller nanoparticles fixed in a matrix and hence have more magnetic moment generated per particle as compared to SHS30. Thirdly and lastly, the use of higher excitation frequency, $f_H$ is favorable for the surface-based MPS bioassay systems.

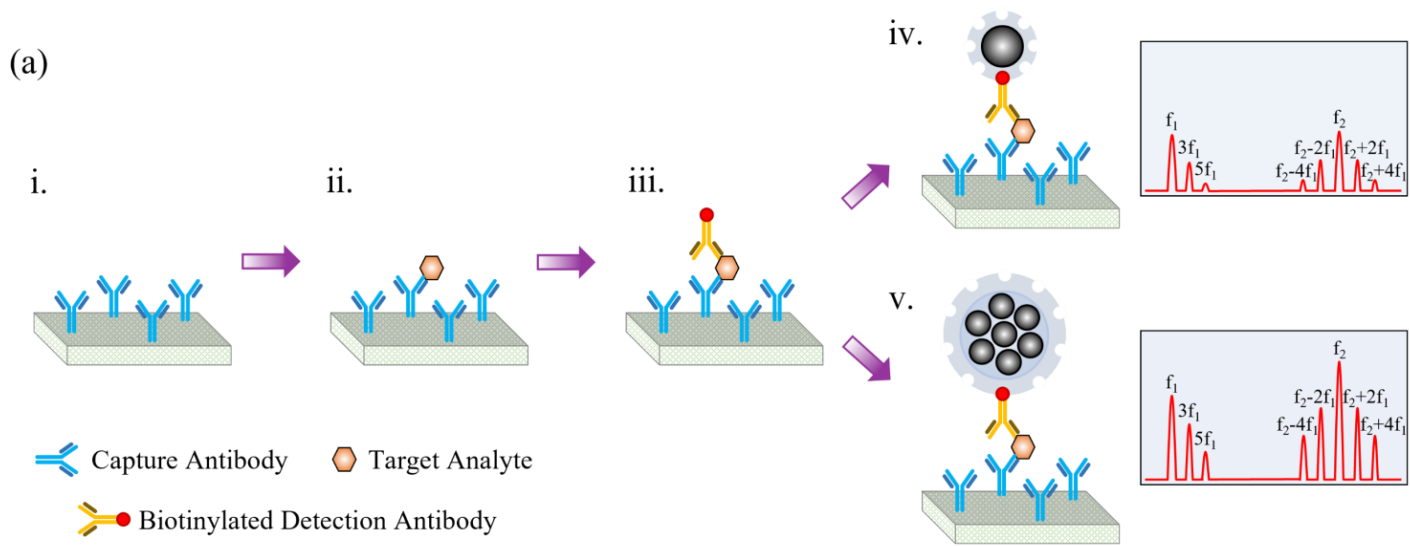

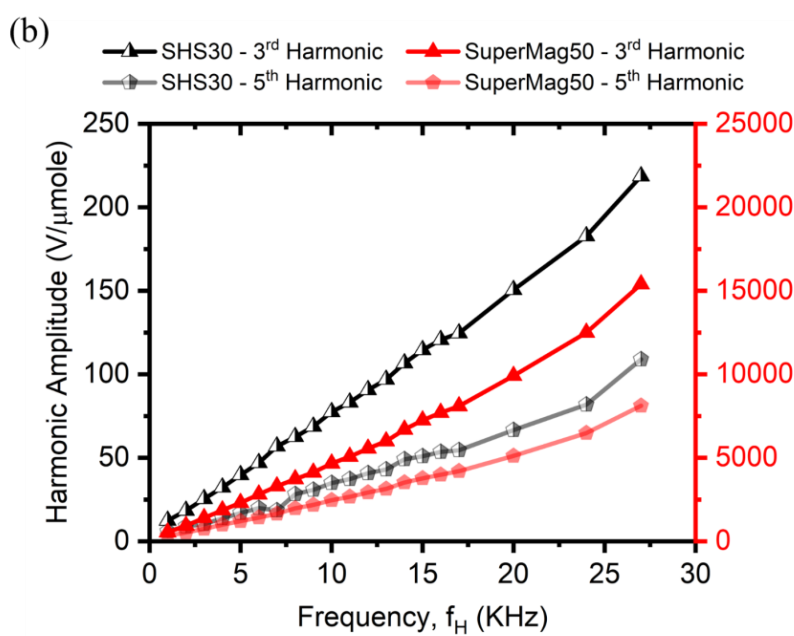

Figure 9. (a) Schematic representation of a surface-conjugation-based MPS bioassay mechanism along with the harmonic signal generation for single- and multicore nanoparticles has been depicted. (i) Capture antibodies are immobilized on a reaction surface. (ii) In addition to the biological test sample constituting a target analyte, the target analyte specifically binds to the capture antibodies. (iii) Biotinylated detection antibodies are immobilized on the target analyte completing the sandwich assay structure. Streptavidin-coated nanoparticles are then captured on the biotinylated detection antibodies, this process for single- and multicore MNPs has been shown in (iv) and (v) respectively with corresponding resultant MPS harmonic spectra; (b) Harmonic amplitude per micro-mole of bound-state of MNPs for SHS30 and SuperMag50 has been depicted.

## 4. CONCLUSIONS AND FUTURE PROSPECTS

Herein, we study the effects of excitation field amplitude and frequency modulation on the MPS spectra of single-core SHS30 and multicore SuperMag50 magnetic nanoparticles. The modulation study was performed for both the single- and dual-frequency MPS schemes to uniquely identify the optimization conditions that can improve the bioassay system sensitivities. For these experiments, MNP samples in unbound and bound states were prepared to simulate (1) border scenarios of a volumetric bioassay scheme, and (2) immobilization scenarios of a surface-based bioassay scheme. Streptavidin-coated nanoparticles were added with excess amounts of biotinylated antibodies to cause aggregation and complete sedimentation to realize the bound state. To identify the optimal scenarios for surface-based bioassay applications, the harmonic signal from the bound state of nanoparticles was chosen to be the determining factor; and to note the favorable conditions for volumetric bioassay systems, the higher harmonic signal from an unbound state of MNPs together with higher Δ were chosen to be the deciding factors for optimized performance. The findings were noted as described below.

The experimental results demonstrate the distinct advantages of choosing multicore MNPs for use with dual-excitation frequency MPS systems in surface-based bioassays. We further noted that for this modality, modulating the excitation field frequency and amplitude could further improve the system sensitivities. Firstly, for the optimal frequency selection, a higher excitation frequency was found to be favorable as it generated a higher harmonic signal from the same MNP amount. Secondly, for the optimal excitation field amplitude selection, it was found that for SuperMag50 multicore MNPs, $A_L = 62.5$ Oe and $A_H = 25$ Oe combination resulted in the highest harmonic signal from the bound state of nanoparticles.

For the volumetric bioassays, both single- and multi-core MNPs in conjunction with the use of a single-excitation frequency MPS system were found to be suitable candidates. As can be noted from Figures 4 and 5, the single frequency systems resulted in a significantly higher Δ whilst providing a better (in case of single core MNPs) or equivalent (in case of multicore MNPs) harmonic spectra amplitudes from the unbound state of nanoparticles when compared to the dual-excitation MPS systems. Furthermore, it was found that for the single excitation frequency MPS scheme, the use of both a higher excitation frequency $f_L$, and larger excitation field strength $A_L$, provide significant improvements in MNP harmonic spectra amplitude and the resultant harmonic drop (Δ). Both favor the scenarios for sensitivity improvement in the case of volumetric bioassays. In this study, however, we do not comment on the choice of single- or multicore MNPs being the better candidate for the volumetric bioassay application. On one hand, single-core MNPs present with a larger harmonic drop meaning that the Brownian relaxation mechanism is more dominant for these nanoparticles, they also offer a smaller harmonic amplitude. This means that a smaller number of multicore MNPs can be utilized instead for the generation of a comfortable harmonic amplitude for bioassay applications which has been shown to provide significant sensitivity improvements[34,35]. Due to this dichotomy between the harmonic amplitude strength and corresponding Δ, the choice between single- or multicore MNPs for better sensitivity has been left for further in-depth investigation.

## ASSOCIATED CONTENT

**Supporting Information**

The supporting information is available at:

- Static magnetization response curves for SHS30 and SuperMag50 MNPs; SPICE modeling of the resonance effects to enable high-frequency high-magnetic field operation; and Dynamic magnetic property characterization using a single-frequency MPS system.

## AUTHOR INFORMATION

**Corresponding Author**

*E-mail: wuxx0803@umn.edu (K. W.)

*E-mail: cheeran@umn.edu (M. C-J. C)

*E-mail: jpwang@umn.edu (J.-P. W.)

**ORCID**

Vinit Kumar Chugh: 0000-0001-7818-7811

Arturo di Girolamo: 0000-0002-6906-8754

Venkatramana D. Krishna: 0000-0002-1980-5525

Kai Wu: 0000-0002-9444-6112

Maxim C-J Cheeran: 0000-0002-5331-4746

Jian-Ping Wang: 0000-0003-2815-6624

**Notes**

The authors declare no conflict of interest.

## ACKNOWLEDGMENTS

This study was financially supported by the Institute of Engineering in Medicine, the Robert F. Hartmann Endowed Chair professorship, the University of Minnesota Medical School, and the University of Minnesota Physicians and Fairview Health Services through COVID-19 Rapid Response Grant. This study was also financially supported by the U.S. Department of Agriculture - National Institute of Food and Agriculture (NIFA) under Award Number 2020-67021-31956. Research reported in this publication was supported by the National Institute of Dental & Craniofacial Research of the National Institutes of Health under Award Number R42DE030832. The content is solely the responsibility of the authors and does not necessarily represent the official views of the National Institutes of Health. Portions of this work were conducted in the Minnesota Nano

Center, which is supported by the National Science Foundation through the National Nano Coordinated Infrastructure Network (NNCI) under Award Number ECCS-1542202.

# Supporting Information

# Frequency and Amplitude Optimizations for Magnetic Particle Spectroscopy Applications

Vinit Kumar Chugh[a], Arturo di Girolamo[a], Venkatramana D. Krishna[b], Kai Wu[a,#,*], Maxim C-J Cheeran[b,*], and Jian-Ping Wang[a,*]

[a]Department of Electrical and Computer Engineering, University of Minnesota, Minneapolis, MN 55455, United States

[b]Department of Veterinary Population Medicine, University of Minnesota, St. Paul, MN 55108, United States

*E-mails: kai.wu@ttu.edu (K. W.), cheeran@umn.edu (M. C-J. C.), and jpwang@umn.edu (J.-P. W.)

#*current address:* Department of Electrical and Computer Engineering, Texas Tech University, Lubbock, TX 79409, United States

## S1. Static (dc) magnetization response curves for SHS30 and SuperMag50 MNPs

Characterization for the static magnetic properties of SHS30 and SuperMag50 was done using the physical property measurement (PPMS) system by Quantum Design. The MNP samples were modulated under varying magnetic field in the range ±500 Oe to obtain the information regarding the coercive fields ($H_c$), and in the range ±5000 Oe to gauge the saturation magnetization ($M_s$) values. Measurements were performed at the temperature of 300 K, having field incremental steps of 10 Oe and an averaging time of 100 msec per step. The dc M-H curves thus obtained have been presented in Fig. S1. From the measurements, observed Ms values for SHS30 and SuperMag50 were 63.8 emu/g and 55.1 emu/g respectively. For the coercive fields, SHS30 showed a recorded value of 17.1 Oe whereas the SuperMag50 sample did not show any coercivity for the measurements performed at ±500 Oe. The minimal coercivity observed for SHS30 MNPs did not affect the MNP suspension stability due to presence of surface proteins which in turn increase the interparticle distance. Suspension of both the MNPs were found to be stable for up to a week.

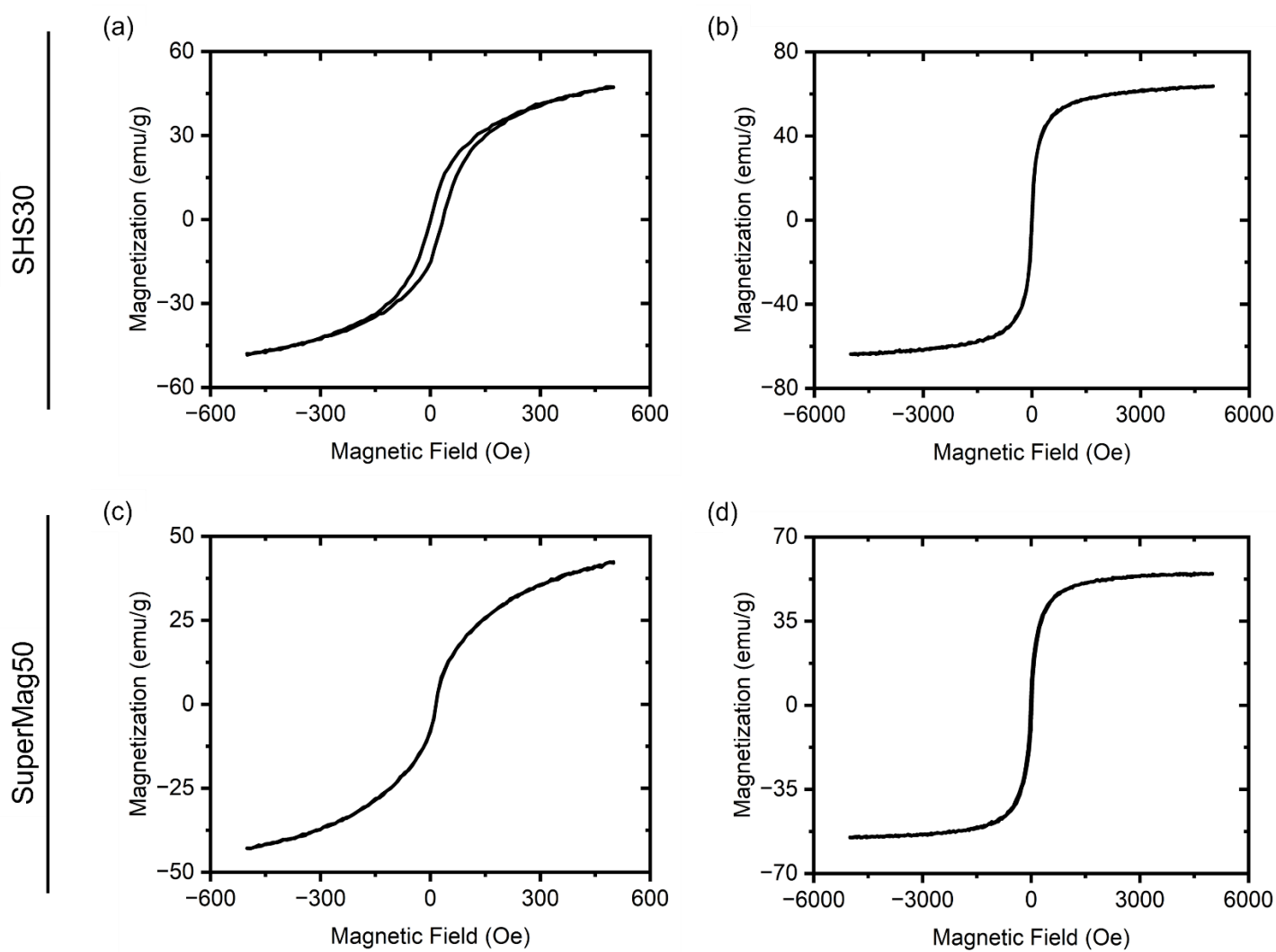

Fig. S1. Magnetization curves for SHS30 and SuperMag50 MNPs with (a) 500 Oe applied magnetic field, and (b) 5000 Oe applied magnetic field.

**S2. SPICE modeling of the resonance effects to enable high-frequency high-magnetic field operation**

SPICE modeling showing how resonating capacitors enable use of high frequencies for both low-frequency ($f_L$) and high-frequency ($f_H$) excitation fields has been presented. The simulations were performed using the LTspice simulation tool to study the current waveform through coil elements whilst a constant voltage is maintained. The coil parameters used for $f_L$ (Primary coil) and $f_H$ (Secondary) coils, for the simulation purposes, have been presented in Table S1.

| **Primary Coil** | | **Secondary Coil** | |
|---|---|---|---|
| Resistance, $R_P$ | 7.923 Ω | Resistance, $R_S$ | 7.878 Ω |
| Inductance, $L_P$ | 14.94 mH | Inductance, $L_S$ | 694.7 µH |
| Capacitance, $C_P$ | 1.36 nF | Capacitance, $C_S$ | 1.36 pF |

Table S1. Coil parameters for the low-frequency (Primary) and the high-frequency (Secondary) coils used for resonance simulation purposes.

For first creating a coil resonance structure and then to be able to shift the resonant frequency in the region of interest, a series resonant capacitor, $C_R$ was added as depicted in Fig. S2. The schematic diagram of the components used for resonance simulations for Primary and the Secondary coils have been presented in Fig. S2(a) and (c) respectively. The resonance capacitor values for the Primary coil circuit were varied between no-capacitor, 10 µF, and 200 nF. For the Secondary coil, the chosen resonant capacitor values were no-capacitor, 1µF, and 20 nF. The Primary coil was excited by a 12V sinusoidal signal whereas a 10V ac voltage source was used for the Secondary coil to perform the resonance analysis. With these modulations in place, ac analysis was performed to plot current through the coils as the frequency of voltage source is varied with corresponding results being plotted in Fig. S2 (b) and (d) for Primary and Secondary coils respectively. A shift in the resonant peaks to higher frequency region can be observed on use of a smaller capacitance value with no deterministic change in the current peak and hence showing the feasibility of using higher frequency excitation fields with resonance implementation has been depicted.

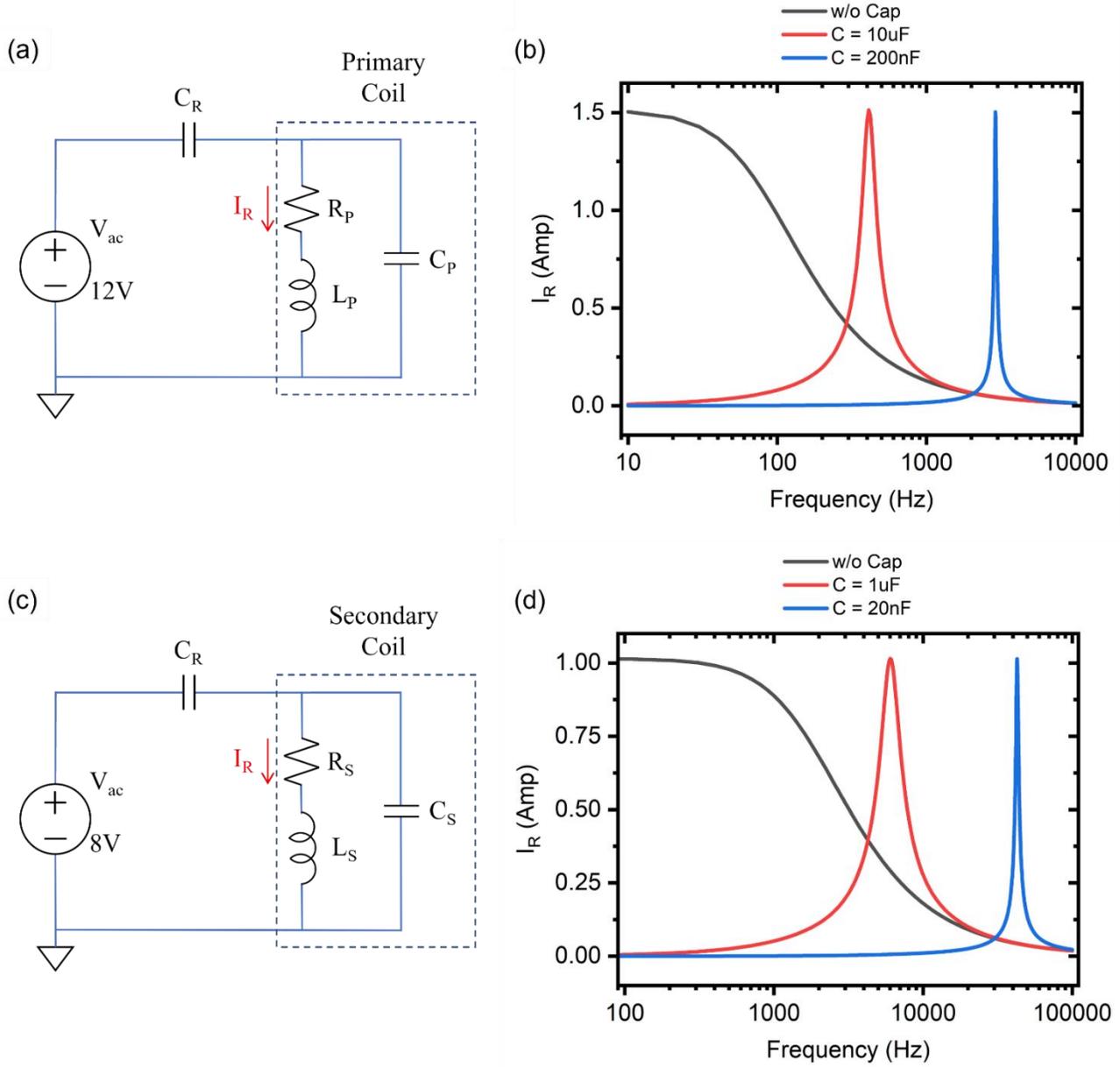

Fig. S2. Schematic view of the component arrangement for resonance simulations for the Primary and the Secondary coils have been presented in (a) and (c) respectively. The current waveforms from ac analysis performed for different resonant capacitance values have been presented in (b) and (d) for the Primary and the Secondary coils respectively.

## S3. Dynamic magnetic property characterization using a single-frequency MPS system

The dynamic magnetic property characterization of the MNPs can be achieved by use of single-frequency MPS modality. In this section, we will try to address the two main points namely, (1) Why do the dynamic magnetic characterization; and (2) How is it achieved with the voltage waveform collected from the MPS systems. The static magnetization response curve obtained through VSM, uses static DC field increments of the applied field to draw the magnetization response curve. This phenomenon allows a MNP magnetization to follow and settle to its maximum value at any given applied field, which on one hand is useful to get the critical information like saturation magnetization ($M_s$) but lacks the information regarding the phase lag response of MNPs to any applied magnetic field. Hence, the AC magnetization response finds its use in determining the magnetization phase lag information when applied with an AC excitation field.

Fig. S3 shows the systematic progression for drawing the AC magnetization loops from the voltage response captured in a MPS system. The pickup coil generates a voltage corresponding to the magnetization response of MNPs when an AC magnetic field is applied as depicted in Fig. S3(a). This voltage generation follows the Faraday's law of induction

$$\varepsilon = -N\frac{\Delta\phi}{\Delta t}$$

Where magnetic flux, $\phi = m \times Area$ ($m$ = magnetization response of the MNPs). Hence, we can back calculate the normalized magnetization, $m$ from the voltage response, $\varepsilon$ by integrating over the one period of applied magnetic field. The normalized magnetization response thus calculated is also presented in Fig. S3(a). Now with utilizing the applied field information together with the magnetization response information, AC magnetization response curve can be drawn as presented in Fig. S3(b).

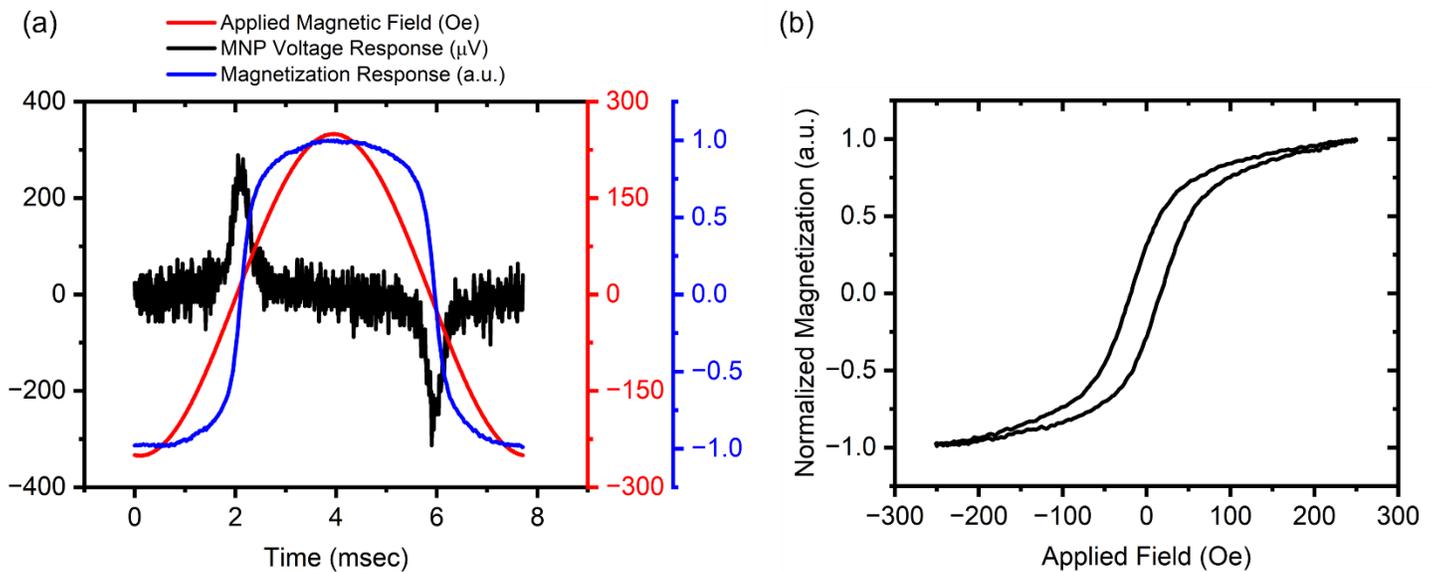

Fig. S3. Time domain signals corresponding to the single tone applied magnetic field, MNP voltage response, and evaluated magnetization response have been presented for SHS30 MNPs in (a). The calculated AC M-H loop from the applied field and the resulting magnetization has been depicted in (b).